\documentclass[printer]{aa} 

\usepackage{natbib}
\usepackage{graphicx}
\usepackage{xspace}
\usepackage{latexsym}
\usepackage{amssymb}
\usepackage{amsmath}
\usepackage{epsfig}
\usepackage{hyperref}
\usepackage{subfigure}
\usepackage{color}
\usepackage{wrapfig}
\usepackage[varg]{txfonts}
 \bibpunct{(}{)}{;}{a}{}{,} 

\usepackage{graphicx}
\usepackage{txfonts}
\begin{document}

 \title{Simulating charge transport to understand the spectral response of Swept Charge Devices}


   \author{P. S. Athiray
          \inst{1,2},
          P. Sreekumar\inst{2},
          S. Narendranath \inst{3},
          \and
          J. P. D. Gow \inst{4}
          }
   \institute{Manipal Centre for Natural Sciences (MCNS), Manipal University, Manipal, India
   \and Indian Institute of Astrophysics (IIA), Koramangala, Bangalore,
              India
              \email{athray@gmail.com}
             \and ISRO Satellite Centre, Bangalore, India
             \and Centre for Electronic Imaging, The Open University, UK}


  \abstract
  {Swept Charge Devices (SCD) are novel X-ray detectors optimized for
      improved spectral performance without any demand for active cooling. The
      Chandrayaan-1 X-ray Spectrometer (C1XS) experiment onboard the Chandrayaan-1
      spacecraft used an array of SCDs to map the global surface elemental
      abundances on the Moon using the X-ray fluorescence (XRF) technique.
      The successful demonstration of SCDs in C1XS spurred an enhanced version of
      the spectrometer on Chandrayaan-2 using the next-generation SCD sensors.
      }
  {The objective of this paper is to demonstrate validation of a physical model developed to
   simulate X-ray photon interaction and charge transportation in a SCD.  The
   model helps to understand and identify the origin of individual components
   that collectively contribute to the energy-dependent spectral response of the
   SCD. Furthermore, the model provides completeness to various calibration tasks,
   such as generating spectral response matrices (RMFs - redistribution
   matrix    files), estimating efficiency, optimizing event selection
   logic, and    maximizing event recovery to improve photon-collection
   efficiency in SCDs.
   }
   {Charge generation and transportation in the SCD at different layers related to   channel stops, field zones, and field-free zones due to photon interaction were
   computed using standard drift and diffusion equations.  Charge collected in
   the buried channel due to photon interaction in different volumes of the
   detector was computed by assuming a Gaussian radial profile of the charge
   cloud. The    collected charge was processed further to simulate both diagonal
   clocking read-out,  which is a novel design exclusive for SCDs,    and event
   selection logic to construct the energy spectrum}.
   {We compare simulation results of the SCD CCD54 with measurements obtained
       during the ground
   calibration of C1XS and clearly demonstrate that our model reproduces all
   the major spectral features seen in calibration data. We also
   describe   our understanding of interactions at different layers of SCD that
   contribute to the observed spectrum. Using simulation results, we identify
   the  origin of different spectral features and quantify their contributions.
   }
{}
   \keywords{X-rays:general -- instrumentation:detectors -- methods:numerical}
   \titlerunning{Simulating charge transport in SCD}
   \authorrunning{Athiray et al.,}

   \maketitle

%

\section{Introduction}

Swept Charge Devices (SCD) are one-dimensional X-ray CCDs developed
by e2v technologies Ltd., UK, to achieve good spectral performances at elevated
operating temperatures. An array of twenty-four SCDs (CCD54) with a geometric area of
${\sim}$ 1$cm^2$ each was used in the C1XS experiment \citep{Ho09, Gr09} onboard the Chandrayaan-1
spacecraft.  Chandrayaan-1 was launched successfully on 28 October 2008
with eleven scientific experiments to study the Moon in multiwavelengths. C1XS
had the opportunity to decipher the lunar surface chemistry by measuring the XRF
signals of all major rock-forming elements, (Na), Mg, Al, Si, Ca, Ti, and Fe,
which are excited by the solar X-rays on the Moon simultaneously.

C1XS is one of the well-calibrated X-ray instruments that studied
the chemical composition of the lunar surface with a good spectral resolution
(onboard resolution ${\sim}$ 153 eV at 5.9 keV). C1XS observed the Moon
under different solar flare conditions for a period of nine months from November
2008 to August 2009. C1XS observations were duly supported by the simultaneous
measurement of incident solar X-ray spectrum by the X-ray Solar Monitor (XSM)
onboard Chandrayaan-1.  Along with detecting major rock-forming elements
\citep{Nar11}, C1XS was the first instrument to observe the direct spectral
signature of the moderately volatile element Na (1.04 keV) \citep{Athiray14}
from the lunar surface. These results opened up a new dimension in our current
understanding of the lunar surface evolution and raised new questions about the
inventory of volatiles on the Moon.  Unfortunately C1XS could not complete its objective of global elemental
mapping due to the lack of solar X-ray activity and
limited mission life. The limited weak flare C1XS observations resulted in a coarse
spatial elemental mapping of certain regions of the Moon.  The upcoming
Chandrayaan-2 Large Area Soft x-ray Spectrometer (CLASS) experiment is being
developed to continue and complete the global mapping of lunar surface
chemistry. CLASS will use the second generation of SCDs (CCD236), which benefit
from increased detector area and modified device architecture for improved
radiation hardness \citep{Gow12}.

X-ray detectors  with good sensitivity and spectral resolution are needed to spectrally resolve the closely spaced lunar XRF
lines  Na (1.04 keV), Mg (1.25 keV), Al (1.48 keV), and Si (1.75 keV).
The measured XRF line fluxes should have minimal uncertainties because the errors in
the derived elemental abundances are directly coupled to XRF line flux errors
\citep{Athiray13}. Thus the accuracy of results depends on a good understanding of
the instrument and the construction of a realistic detector spectral response from calibration data. A  physical model incorporating the physics of charge generation and transportation in a device
whose architecture is known provides complementarity and completeness to the
measured calibration data, while also enabling the tuning of the device operation and
event-selection criteria to optimize performance. In this paper, we present the charge transport model developed to simulate the spectral response of the  SCD.\\

 In Sect. \ref{scdcxs} we describe the architecture of the SCD CCD54 used in C1XS
 and explain its clocking and read-out mechanism along with the different event
 selection processes adopted in the C1XS experiment. The algorithm developed to model charge transport, a description of charge
   transport model with fundamental assumptions, necessary equations, and its implementation are all explained in detail in Sect.\ref{algo}.  A brief summary of the C1XS ground
   calibration is presented in Sect.\ref{gc}, which are used for the validation
   of charge transport model. Salient features of the simulation,
   results, and comparison with C1XS ground calibration data are presented in
   Sect. \ref{smres}, and the conclusions inferred from this model are summarized in Sec. \ref{conc}.

  \section{Swept Charge Devices used in C1XS}
  \label{scdcxs}
  SCDs are modified versions of conventional X-ray CCDs specially designed for
  non-imaging,  spectroscopic studies. The C1XS experiment used an array
  of 24 SCDs \citep{Low01} to record X-ray emission in the energy range of
  0.8 to 20 keV. It offers good spectral resolution at benign operating
  temperatures, allowing the C1XS detectors to be operated between -10$\,^{\circ}$C to +5$\,^{\circ}$C
  using a passively cooled system. The SCD CCD54 has an active area of 1.07
  cm$^2$ per SCD and contains 1725 diagonal silicon electrodes with channel
  stops arranged in a herringbone structure, and the structure is shown in Fig. \ref{ccd54}.
  The device has an n-type buried channel beneath the
  electrodes where charges generated by photon or particle interactions are
  collected. The pitch of the channel stop is 25 ${\mu}$m \citep{jgow09} and require
  575 clock triplets to completely flush the SCD. \\

  \subsection{Clocking and read-out in SCD}
\begin{figure}%
  \centering
  \subfigure[]{
  \includegraphics[height=5.5cm]
  {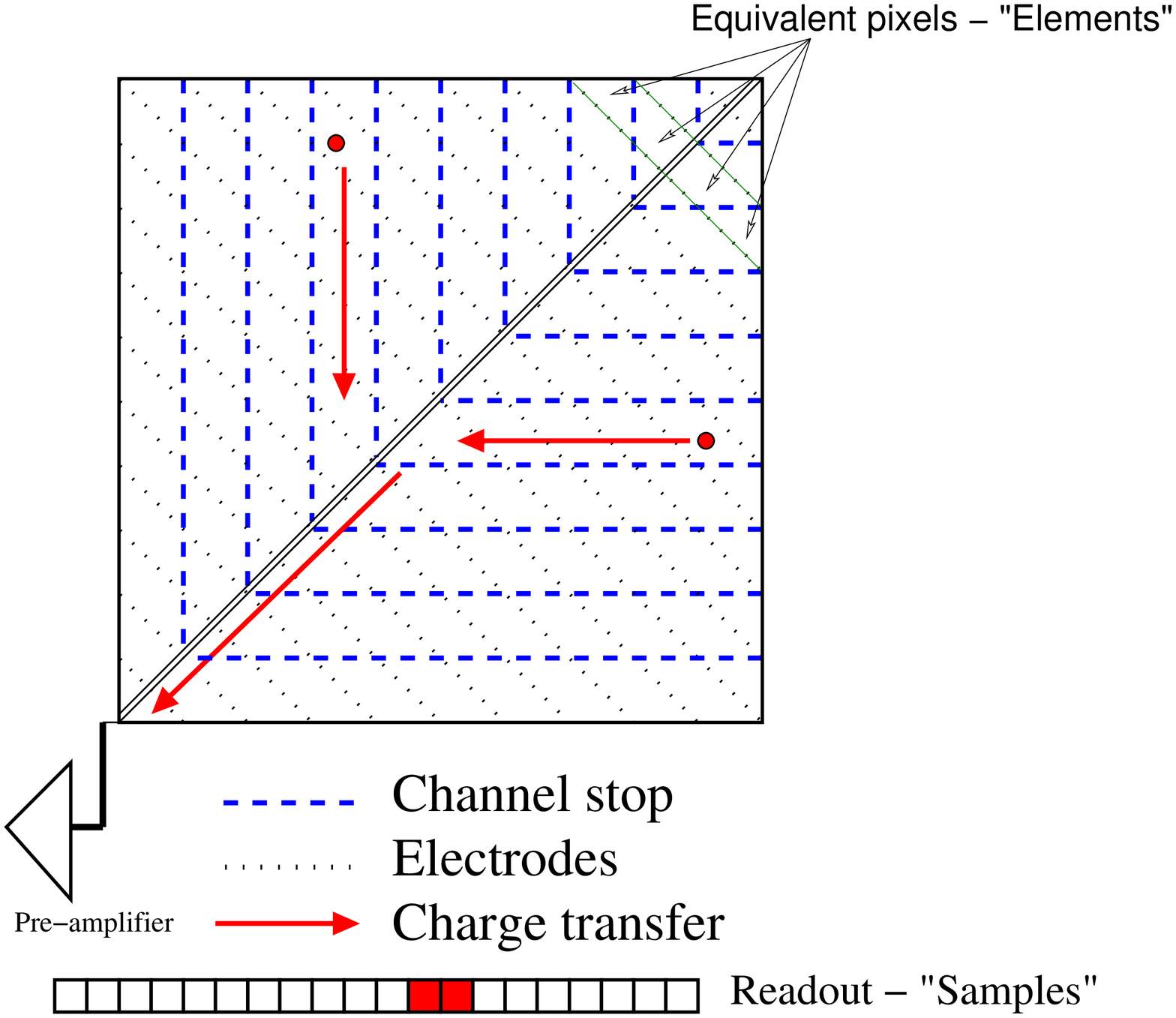}
  \label{ccd54}%
  }%
  \qquad
  \subfigure[]{
  \includegraphics[height=5.5cm]{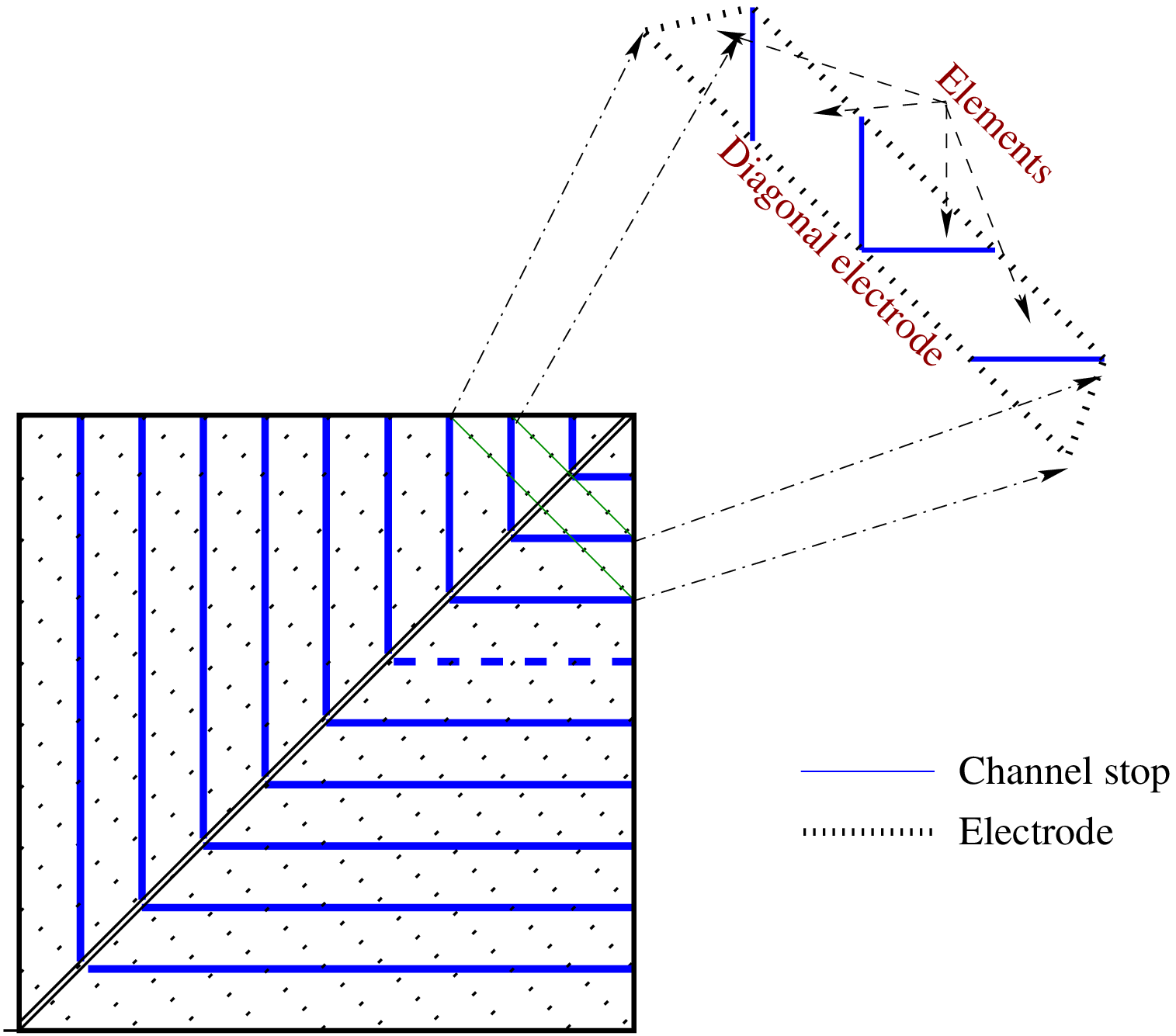}
  \label{ccd54elements}%
  }%
  \caption{(a) Schematic view of the SCD CCD54 (b) Representation of elements in a diagonal electrode of SCD CCD54.
  Regions in the zoomed electrode are the elements which are separated by channel
stops.}
  \label{ccd54arch}
  \end{figure}

  The SCD operation is  similar to a conventional X-ray CCDs where
  clock voltages are applied to electrodes to transfer charges collected in
  the buried channel. The difference is due to the novel electrode architecture implemented in the SCD, shown in Fig. \ref{ccd54}, which enables a large detector area to be read out quickly. SCDs are operated in continuous clocking mode at high frequencies  (${\sim}$100 kHz) \citep{jgow09}, the continuously clocking suppresses the surface component of dark current and allows for operation at warmer temperatures when compared to a conventional 2D X-ray CCD. The
  term "pixel", which refers to photon interaction co-ordinates on the device in
  conventional CCD, is not strictly applicable for a SCD. We refer to the photon
  interaction region on  a device   as {\it elements,} which are shown in Fig.
  \ref{ccd54elements}. These {\it elements} in each electrode are analogous to
  pixels in a 2D   X-ray
  CCD. Charges collected within each {\it element} under different electrodes
  are clocked toward the central diagonal channel   and then clocked down to reach the
  readout node (arrows in Fig. \ref{ccd54} indicate direction of charge flow).
  Charges collected in each electrode are subject to the same  number of clock cycles to reach the readout node where
  they are merged to give a   pseudo linear output. We refer to the
  combination of charges from these {\it elements} at the readout node as {\it
  samples}.  The result of this method of operation is that the linear readout does not immediately provide information about where the photons
  were incident. Various advantages of SCD over conventional X-ray
  CCDs are given in \cite{jgow09}.

  \subsection{Event processing in SCD}
  \label{ep}
  To control data volume, two event-processing modes were adopted in
  C1XS: time-tagged mode (during low and moderate event rates) and spectral
  mode (high event rate) \citep{Ho09}. Because the observed C1XS event rate was low
  throughout the mission, events were processed in time-tagged mode where each
  event was attached with onboard time and SCD number. This mode was further
  subdivided into two other modes based on event rates and a two-threshold logic:
  \begin{enumerate}
  \item{{\bf Multi-pixel mode (${\leq}$ 51 events/s - Type 11 data)}: Out of a group
  of three adjacent pixels, if central pixel is above threshold 1, then store all
  the three events.}
  \item{{\bf Single-pixel mode (51-129 events/s - Type 10 data)} : Out of a
  group of three adjacent pixels, only the central pixel event is stored if it is
  above threshold 1, and both the adjacent pixels are below threshold 2.
  Otherwise discard the central event.}
  \end{enumerate}
  \begin{figure}
  \centering
  \includegraphics[width=8cm]{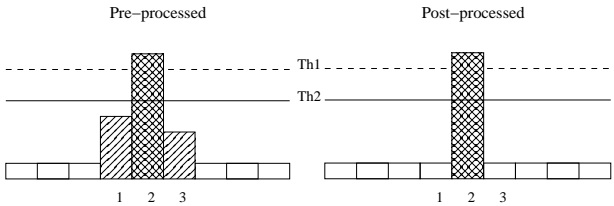}
  \caption{Representation of `Single pixel (Type 10 data)' event selection logic
  used in C1XS with two threshold values {\it Th1, Th2}.}
  \label{eventselection}
  \end{figure}

  Pictorial representation of single-pixel mode is shown in Fig.
  \ref{eventselection}, thus the distribution of charges collected due to a
  photon event varies with different selection modes. This leads to different
  spectral shapes and in turn has an effect on the overall throughput of the detector.

\subsection{Spectral redistribution function (SRF) of SCD}
The interaction of X-ray photons within a detector results in a complex cascade of energy transfers with the energy deposited
by each photon gets transformed in many ways, leading to various signatures in the
observed energy spectrum. The distribution of all energy deposits above a noise
threshold observed in a device by mono-energetic photons is called the spectral redistribution function (SRF).

In a SCD, charges collected from photon interaction are clocked diagonally and
read out as a linear array of charges that are further processed for event
recognition to obtain the spectrum. The observed SRF in SCD is thus a function
of energy of the incident photon, the event selection logic, and threshold
values. This model can be used to optimize threshold values used to select or reject split events. It can also identify the genesis of different
features seen in the observed SRF and enables quantification of different
process components.

  \section{Computation of SRF of SCD}
  \label{algo}
\begin{figure}
  \centering
  \includegraphics[width=9.5cm]
  {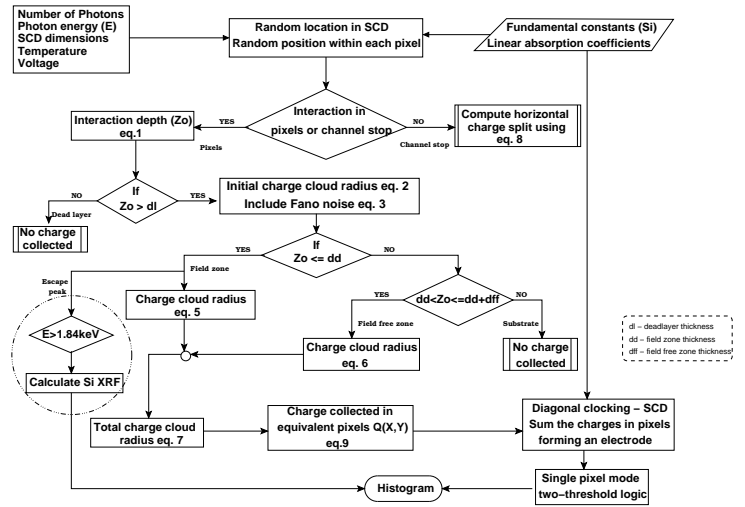}
  \caption{Flowchart of the charge transport model to simulate the SRF of the
  SCD. Some of the important steps involved in the model are arranged in
  sequence. It is clear that the dead layer and substrate are included to
  account for photon loss while no charge is collected from the interaction.}
  \label{ctmalgorithm}
  \end{figure}

\begin{figure}%
  \centering
  \includegraphics[height=6cm,angle=0]
  {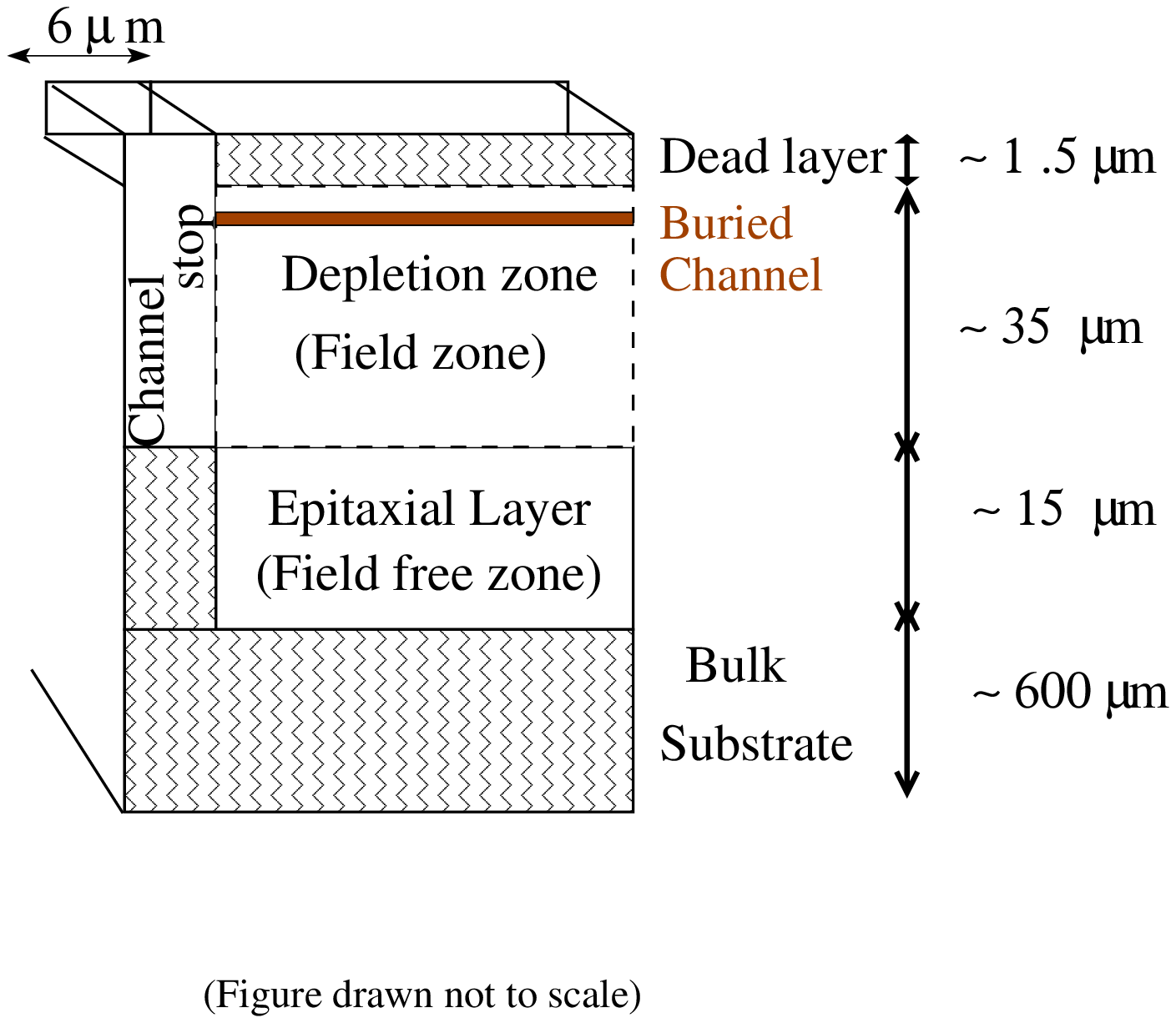}
  \caption{Vertical structure of the SCD CCD54. Photon interaction in field
  zone, field-free zone, and channel stops contribute chiefly to the observed
  SRF. Interaction in bulk substrate and regions in dead layer and above result in
  a negligible amount of final charge collection.}
  \label{ccd54_structure}%
  \end{figure}

  In this section we describe the Monte Carlo algorithm used to simulate X-ray
  photon interactions and charge propagation in the SCD. A flowchart
  explaining the steps involved in the model is shown in Fig. \ref{ctmalgorithm}.
  Fundamental assumptions of the model are listed below:
  \begin{itemize}
  \item Si-based X-ray detector is assumed to be ideal, free of interstitial
  defects and impurities.
  \item The electric field is assumed to only be present perpendicular to the plane of the detector.
  \item The acceptor impurity concentration in the field-free zone is assumed to
  be same as in the field zone.
  \item The boundary between field and field-free zone is modeled with a small
  offset to avoid numerical divergence (at $z_0$ = $d_d$ refer Eq.
  (\ref{fzoner})).
  \end{itemize}
These assumptions are related to the fabrication of X-ray devices for which
direct experimental measurements are difficult to obtain, so major deviations in
these assumptions will affect the detectors' spectral performance. The epitaxial
resistivity only decreases around a few ${\mu}$m near to the field-free zone and substrate boundary interface, which is difficult to measure. Large gradient in epitaxial resistivity amounts to charge losses via recombination. Si with large imperfections and defects will contribute to enhanced dark noise resulting in spectral degradation. However, no such effects were experienced in the  ground calibration tests, as well as in flight observations; as a result, these assumptions are considered to be valid.\\
  The vertical structure of the SCD discussed in this paper is shown in Fig.
  \ref{ccd54_structure} with different layers labeled with respective dimensions.
  The dead layer ${\sim}$ 1.5 ${\mu}$m thick, consists of SiO$_2$, Si$_3$N$_4$, and
  poly silicon layer. A negligible fraction of X-ray photons interact in the
  dead layer and result in charge collection. We therefore modeled it as a single
  block of SiO$_2$ from which no charge is collected. The bulk substrate
  is a heavily doped p+ region, where charges suffer huge losses owing to
  recombination, hence are considered to be lost.

  \subsection{Photon interaction}
  Mono-energetic X-ray photons are simulated to illuminate the SCD at random
  positions (x, y) with uniform probability, at normal incidence angle with
  respect to the plane of SCD. These photons travel through different layers of the detector
  before interaction. The distribution of depths at which interactions occur
  inside the SCD is computed using
  \begin{equation}
  \label{eq:depth}
  z_0 = − \frac{1}{{\mu}(E)} ln(R_u)
  \end{equation}
  where the ${\mu}$(E) - linear mass absorption coefficient of the material at
  photon energy E, R$_u$ - random number with uniform distribution.  If the interaction depth
  (z$_0$) is greater than the thickness of a layer, then the interaction depth is
  computed again, including the material in the following layer. Soft X-ray photons
  interact via the photoelectric process, resulting in a charge cloud (e-h pairs), which is
  assumed to be spherical in shape.  The radial charge distribution of this
  spherical cloud is assumed to follow a Gaussian distribution with the 1${\sigma}$ radius given by \cite{Kurniawan07}
   \begin{equation}
  \label{eq:initialchargecloudradius}
  r_i =\left\{ \begin{array}{ll}
  40.0 \frac{E_{pe}^ {1.75}}{{\rho}} & (5~keV < E_{pe} < 25~keV)\\
  & \\
  30.9 \frac{E_{pe}^ {1.53}}{{\rho}} & (E_{pe} {\leq} 5~keV)
  \end{array}
  \right.
  \end{equation}
  where ${\rho}$ - density of the detector material (${\rho}$ = 1.86 g/cc for Si),
  E$_{pe}$ - energy of the photo-electron. This implies that charges are
  concentrated toward the center of the sphere, fall off with radius, and
  abruptly truncate to zero at the cloud boundary. While estimating the number of charges produced
  by absorption of X-ray photon, Fano noise \citep{Fano47, owen02} is
  important,  which is
  incorporated as
  \begin{equation}
  \label{eq:fanonoise}
  E_f = E_i + R_n(0)\sqrt{F{\omega}E_i}
  \end{equation}

  \noindent where $E_f$ - energy distribution with Fano noise added, $R_n(0)$ -
  normally distributed random number with mean 0 and variance 1, F - Fano factor
  (F = 0.12 for Si) and ${\omega}$ - average energy required to produce an e-h
  pair (${\omega}$ = 3.65 eV at 240K for Si). The initial charge ({\it Q$_0$}) is
  obtained from this final energy   distribution (E$_f$) as
  \begin{equation}
  \label{eq:initialcharge}
  Q_0 = \frac{E_f}{{\omega}}
  .\end{equation}

  The interaction of X-ray photons in SCDs can be grouped into three regions based  on
  the photon interaction zone: field zone, field-free zone, and channel stop. The escape
  peak appears as a consequence of X-ray photon interaction within the detector that also needs to
  be incorporated.
  \subsection{Field zone interactions}
  Photons interacting at depths within the depletion zone (i.e., z$_0$
  $<$ d$_d$) are called field zone interactions. The thickness of the
  field zone mainly depends on the bias voltage and doping concentration.
  In the case of the SCD CCD54, with an acceptor impurity concentration (N$_a$) of
  4${\times}$10$^{12}$ atoms/cm$^3$ and average depletion voltage of 9 V (\cite{jgow09},
  the depletion depth was found to be ${\sim}$ 35 ${\mu}$m. The charge cloud produced
  here by a photon interaction will experience the
  complete electric field and will drift toward the buried channel. The radius of
  the charge cloud collected within the buried channel after charge spreading due to
  random thermal motions is given by \citep{Hop87}
  \begin{eqnarray}
  \label{fzoner}
  r_d = \sqrt{4Dt} = \sqrt{\frac{4KT{\epsilon}}{e^2N_a}~~ln~(\frac{d_d}{z_d-z})}
  \end{eqnarray}
  where K  is the Boltzmann constant, T the temperature in kelvin, e the electronic
  charge, ${\epsilon}$ the permitivity of Si, N$_a$
  the  doping concentration in Si, and d$_d$ the
  thickness of depletion depth.
  \subsection{Field-free zone interactions}
  In epitaxial devices, there is a field-free region between the field zone and
  the bulk Si substrate (p+) layer, which does not have an electric field (see
  Fig. \ref{ccd54_structure}) and exhibits a gradient in doping concentration
  near to the substrate boundary. Photon interactions in this region $d_d < z_0
  < d_d+z_{ff}$ are considered to be field-free zone interactions.
  Charge clouds produced within this region will not experience any acceleration but could
  diffuse and recombine before reaching the buried channel. Diffusion enlarges
  the size of the charge cloud and could cause spreading of charges across
  several adjacent elements. It is reasonable to assume that the recombination losses are negligible in this region due to large carrier
  lifetime (${\sim}$ 10$^{-3}$ s) \citep{tyagi83}. The general expression for the
  mean-square radius of charge cloud reaching the interface between the field
  and field-free zone is given by \citep{pav99}
  \begin{equation}
  r_{ff} = \sqrt{2d_{ff}L~\biggl[tanh(\frac{d_{ff}}{L}) -
  (1-\frac{z-d_d}{d_{ff}}) tanh\biggl(\frac{d_{ff}-z+d_d}{L}\biggr)\biggr]}
  \end{equation}
  where $d_{ff}$ is the thickness of epitaxial field-free zone, L = diffusion length
  (L = $\sqrt{D{\tau}_n}$, ${\tau}_n$ is the charge carrier lifetime, and D the  diffusion constant). It is evident from the above equation that when a photon
  interacts deeper in the field-free zone, the term ($d_{ff}-z_0+d_d$) becomes
  very very small, so the radius ($r_{ff}$) becomes very large. It should be
  noted that the shape of the charge cloud in the field-free zone is
  non-Gaussian as investigated by \cite{pav99}. However, here we assume that they are Gaussian because it only affects a few  high energy photon
  events in the field-free zone. Thus, the final radius of the charge cloud is
  obtained as the quadrature sum of $r_{i}$, $r_{d}$, $r_{ff}$ :

  \begin{equation}
  r = \sqrt{r_i^2+r_d^2+r_{ff}^2}
  .\end{equation}
  \subsection{Channel stop interactions}
  The channel stops occupy a considerable amount of area in the CCD54 (${\sim}$ 20\%), and photon interaction in channel stops can distort the shape of the observed SRF because it
  causes horizontal split events. We adopted the same approach followed in
  simulating the response of the CCDs in the Chandra Advanced CCD Imaging Spectrometer (ACIS) \citep{town02}, where photons interacting
  in the channel stops (p+ layer) are assumed to split the resulting charge
  and are collected on the left- and the righthand sides of the channel stop. The fraction of
  charges swept toward each side depends on the distance of each side from the
  location of interaction. Charges propagated from both the edges of a channel
  stop are found using \citep{town02}
  \begin{equation}
  \label{eq:channelstopcharges}
  \begin{array}{ll}
  Q_{left} =\frac{w_c - x_c}{w_c} Q_0 - {\chi}x_c + R_N(0){\alpha}\frac{x_c}{w_c}\\
  & \\
  Q_{right} =\frac{x_c}{w_c} Q_0 - (w_c - x_c){\chi} + R_N(0){\alpha}\frac{w_c - x_c}{w_c}
  \end{array}
  \end{equation}
  where w$_c$ is the width of channel stop, ${\chi}$ the channel stop tuning
  loss parameter, and ${\alpha}$ the channel stop tuning width parameter. The interaction in within the channel stop region is complex because it involves interaction in the p+ layer or
  directly underneath. It is shown from the detailed ACIS calibration
  \citep{town02} that the channel-stop tuning parameters ${\chi}$ and ${\alpha}$ vary with
  incident photon energy, which require precise experimental data on channel
  stop interactions.  However, for simplicity, we assume energy independence of
  these parameters in this work. From Eq. \ref{eq:channelstopcharges} it is
  clear that a change in ${\chi}$ and ${\alpha}$ with photon energy will modify the fraction of
  charges collected on either side of the channel stop and is expected to
  affect the distribution of horizontally split events, and this contributes to the
  non-photopeak events of the SRF.
 \subsection{Charge collection and read-out}
  The   amount of charge collected in  each element (i, j) at the gates is
  obtained by integrating the Gaussian function, which is given by  \cite{pav99}:
  \begin{equation*}
  Q_{ij}(x_0,y_0) = {\int}_{a_{i}}^{a_{i+1}}{\int}_{b_{i}}^{b_{i+1}} q({\bf r})~dxdy
  \end{equation*}

  \begin{equation}
  \label{eq:chargecollected}
  Q_{ij}(x_0,y_0) = \frac{Q_o}{4} \begin{array}{ll}
  \biggl\{\biggl[\mbox{erf}\biggl\{\frac{a_{i+1}-x_0}{r}\biggr) - \mbox{erf}\biggl(\frac{a_i-x_0}{r}\biggr)\biggr]
  \biggl[\mbox{erf}\biggl(\frac{b_{i+1}-y_0}{r}\biggr)  -
  \mbox{erf}\biggl(\frac{b_i-y_0}{r}\biggr)\biggr]\biggr\}
  \end{array}
  \end{equation}
  where r will be replaced with Eq. (7) for interaction in the field zone and
  field-free zone, respectively; a, b are pixel dimensions; $x_0$ and $y_0$ are the
  relative event
  coordinates in the reference frame with its origin at the center of the pixel
  where the photon is absorbed (-a/2 $< x_0 <$ a/2, -b/2 $< y_0 <$ b/2).\\

  In section \ref{scdcxs}, we have indicated that the fundamental difference between the
  SCD and 2D CCD, primarily lie in the diagonal clocking and read-out. In our
  simulation, we modeled the  {\it elements} as 2D pixels of equivalent
  dimensions. Charges collected in each of these equivalent pixels are computed
  using Eq. (9). Moreover, we incorporated the read-out structure by adding the
  charges collected at different {\it elements} of each electrode in the SCD.
  \subsection{Escape peak computation}
  Photons with incident energy E$_{ph} >$ 1.84 keV (i.e., binding energy of Si
  K-atomic shell electrons) have a finite probability of yielding Si K-${\alpha}$ XRF photons of
  energy 1.74 keV. These XRF photons travel some distance before getting
  absorbed in the detector. XRF photons emitted from the top few microns of
  the detector have a high probability of escaping without being detected. In such cases,
  the residual charges were collected to form the escape peak with energy E$_{esc}$ = E
  - 1.84
  keV. To simulate the escape peak, we find the number of Si XRF
  photons produced from a thick Si medium, where the photons are emitted in all possible directions. Directions are assigned using the uniform random number, in which half
  of the photons exiting the surface of the detector are considered to
  yield the escape peak feature.

  \subsection{Implementation}
  The algorithm explained in Fig. \ref{ctmalgorithm} is implemented with a set
  of IDL routines.  Device level parameters, such as channel stop pitch, width, and acceptor
  impurities (N$_A$) associated with the fabrication of SCD CCD54 provided
  by the manufacturer, are documented in \cite{jgow09}. A thorough characterization and
  optimization of voltages carried out for the flight units of SCD CCD54 to
  achieve the expected spectral performance are summarized in \citep{jgow09}.
  Values of some of the important detector parameters used in
  the  model are listed in Table \ref{tab:parameters}. A short summary of the
  implementation, highlighting the important steps are given below:
 \begin{itemize}
  \item Allow photons to impinge on the SCD randomly at normal
  incidence (i.e., 0$\,^{\circ}\mathrm{}$ with respect to detector normal) and
  obtain subpixel position coordinates (X, Y).
  \item Separately tag photons interacting in channel stops and pixel regions.
  \item Simulate physics of photon interaction and charge propagation separately in field zone, field-free zone, and channel stops.
  \item Clock charges collected in the {\it elements} of SCD
  diagonally, then combine and read out as a linear array output.
  \item Apply single-pixel mode event-selection logic with two threshold values.
  \item Make histogram of the event processed output.
  \end{itemize}
\begin{table}[h]
\caption{Values of parameters used in modeling charge transport of SCD}
\label{tab:parameters}
\begin{center}
\begin{tabular}{|l|l|} 
\hline
{\bf  Parameters} & {\bf Values}\\
\hline
\rule[-1ex]{0pt}{3.5ex} Voltage (V$_T$ for $d_d$ computation) \citep{jgow09}& 3.8 V  \\
\rule[-1ex]{0pt}{3.5ex}  Channel stop pitch \citep{jgow09} & 25 ${\mu}$m  \\
\rule[-1ex]{0pt}{3.5ex}  Channel stop width \citep{jgow09} & 6 ${\mu}$m  \\
\rule[-1ex]{0pt}{3.5ex}  Number of acceptor impurities($N_a$) \citep{jgow09} & 4${\times}10^{12}cm^{-3}$  \\
\rule[-1ex]{0pt}{3.5ex}  Field + field-free zone thick \citep{jgow09} & 50 ${\mu}$m  \\
\rule[-1ex]{0pt}{3.5ex}  Life time (${\tau}$) \citep{tyagi83} & 10$^{-3}$ s  \\
\rule[-1ex]{0pt}{3.5ex}  Temperature \citep{sn11} & 263 K  \\
\rule[-1ex]{0pt}{3.5ex}  Number of mono-energetic photons & 5 ${\times}$ 10$^{4}$ \\
\hline
\end{tabular}
\end{center}
\end{table}

 \section{Ground calibration of the SCD in C1XS - An overview}
  \label{gc}

  An extensive ground calibration was carried out for all the detectors in C1XS,
  using a double crystal monochromator in the RESIK X-ray beam-line at
  Rutherford Appleton Laboratory (RAL), UK. Distinct mono-energies were chosen
  for SRF measurements using the following target anodes: Ti (4.51 keV),
  Cr (5.414 keV), Co (6.93 keV), and Cu (8.04 keV).  The
  mono-energetic beam was collimated using a rectangular slit   of 1mm ${\times}$ 2mm
  dimension illuminating only a small portion of the SCD. Temperature dependence of the SRF parameters was also studied across a range from -30$\,^{\circ}\mathrm{}$ C to -10$\,^{\circ}\mathrm{}$ C. These   calibration
  data were further processed for event recognition using a    two-threshold
  logic explained in Sect. \ref{ep}. Prominent spectral   features seen in the
  filtered calibration data are photopeak, low energy   shoulder, low energy
  tail, cut off, escape peak, and low energy rise. These   spectral features were
  modeled empirically and interpolated throughout the   entire energy range to
  create spectral response matrix of the SCD.  It was   also noted that the
  SRF components exhibit an energy dependence. Further   details on the ground
  calibration of C1XS instrument can be found in   \cite{Nar10}.

  \section{Simulation results - SRF components}
  \label{smres}
  Here we summarize our understanding of different spectral components that
  are seen in the observed SRF. From our simulation, we identified the sources of
  origin for the SRF components, which are summarized in Table (\ref{identify}).

%
\begin{table}
\caption{Identification of SRF components and its origin in the SCD - Simulation
results}             
\label{identify}      
\centering                          
\begin{tabular}{c c}        
\hline\hline                 
SRF component  & Interaction zone \\    
\hline                        
Soft shoulder, Low-energy tail &  Channel stop  \\
Low-energy rise, Low-energy tail   & Field-free zone \\
Photopeak, Soft shoulder, Low-energy tail & Field zone\\
Cutoff, Escape peak &\\      
\hline                                   
\end{tabular}
\end{table}
%

  \subsection{SRF components - Field zone}
 The dominant photopeak in the SRF mainly arise from field zone   interactions,
  where charges are collected almost completely. Interactions   that occur at
  {\it element} boundaries and at greater depths near to the boundary  of
  field-free zone cause charge splitting across many {\it elements}. These
  charges are   clocked diagonally and added appropriately to produce the soft-shoulder and
  low-energy tail. Si XRF   photons emitted from the top layer of the field zone
  have a high probability of escaping and   hence contributing to the escape peak.
  The simulated SRF due to field zone interactions for different X-ray energies
  are shown in Figs. \ref{ti}, \ref{cr}, \ref{co} and \ref{cu}.\\ Cutoff is observed at two
  places in the energy spectrum, which are related to   the value of threshold 1
  (Th1):
  \begin{itemize}
  \item Low energy cutoff  : Events below threshold 1; i.e., 0.75 keV are discarded.
  \item Soft shoulder cutoff: Charges that are split at pixel boundaries lead to
  configurations with central event discarded (see Fig.
  \ref{eventselection}) causing a dip near the soft shoulder. The energy at
  which dip occurs in the spectrum depend on threshold 1 given by E$_{dip}$ ${\approx}$ E$_{ph}$ - E$_{th1}$.
  \end{itemize}
  \subsection{SRF components - Field-free zone}   The charge cloud produced due to
  field-free zone interactions undergoes diffusion,   causing charge to spill over
  multiple {\it elements}. As a result, a low-energy   rising-tail component
  without a photopeak is observed in the pulse height   distribution as shown in
  Figs. \ref{ti}, \ref{cr}, \ref{co}, and \ref{cu}. This component is negligible
  for low energy X-ray photons as  a majority of photons get absorbed above the
  field-free zone.
  \subsection{SRF components - Channel stop}
\begin{figure}%
\centering
\subfigure[]{
        \includegraphics[height=6cm]
                {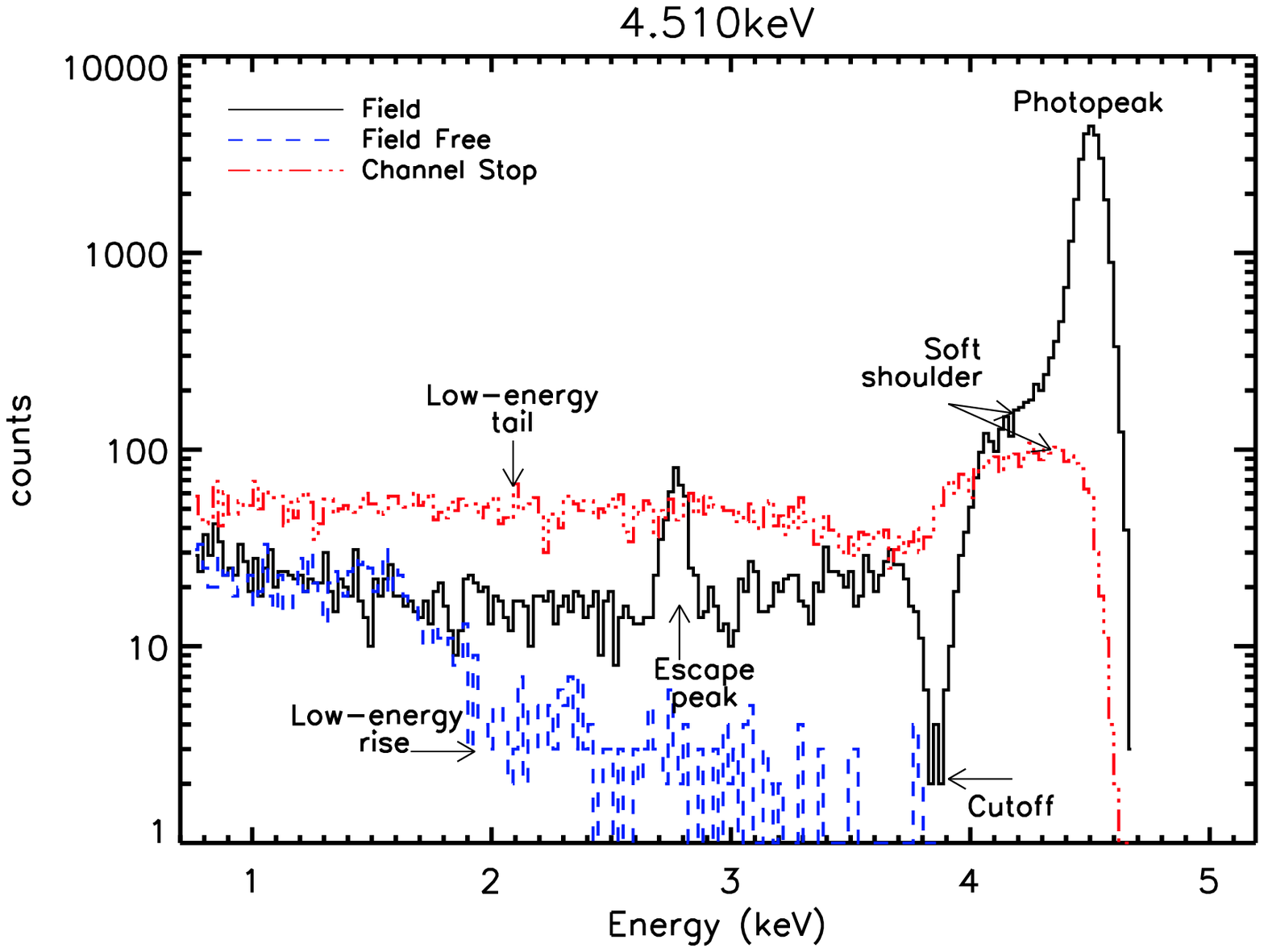}
\label{ti}%
}%
\qquad
\subfigure[]{
        \includegraphics[height=6cm]
                {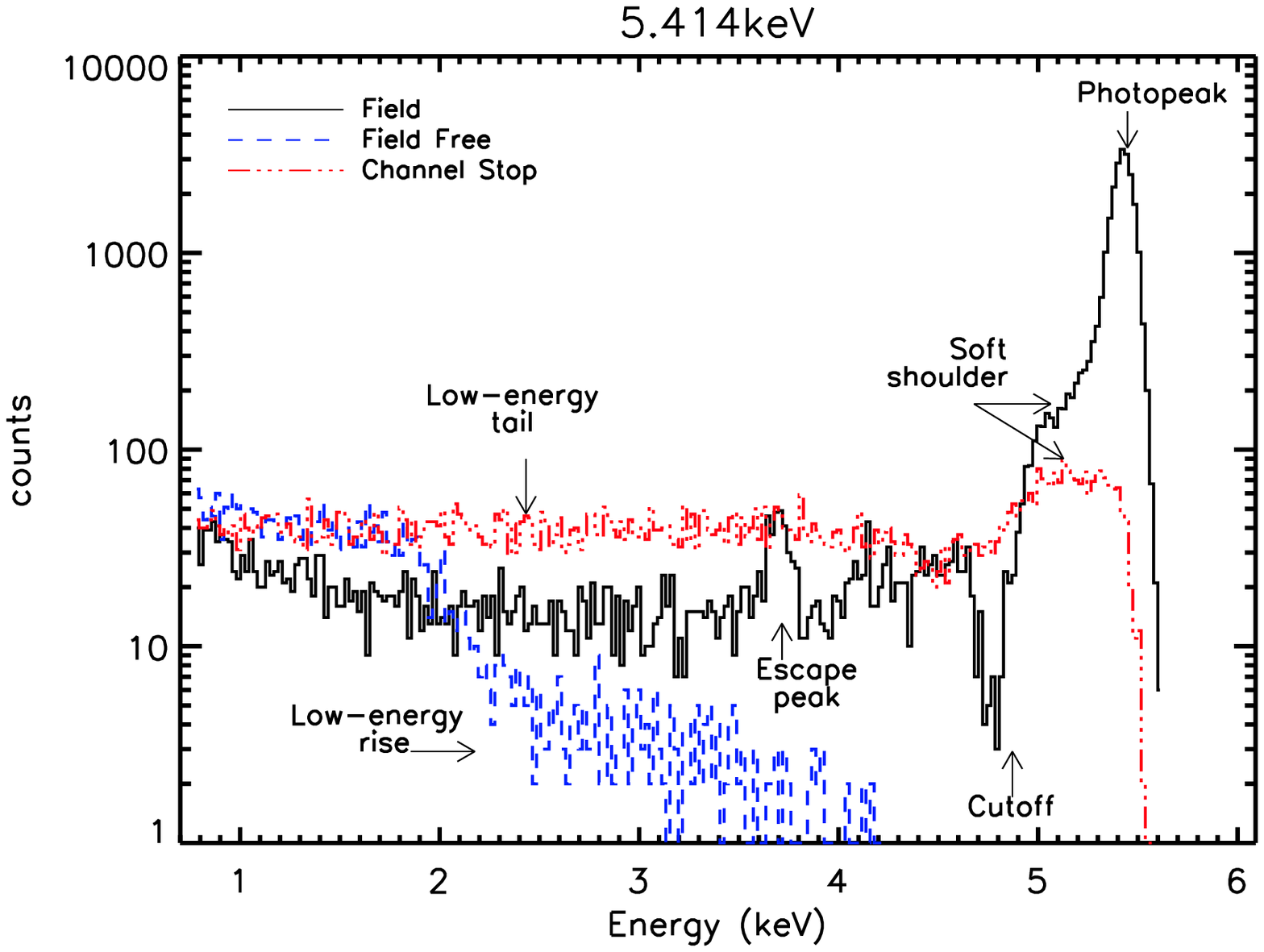}
\label{cr}%
}%
\caption[SRF of SCD with components - simulation results (4.510 keV, 5.414 keV)]{Simulation results showing different components of SRF for
different mono-energetic photons arising from interactions at different
layers of SCD: (a) 4.510 keV (b) 5.414 keV.}
\end{figure}

\begin{figure}%
\centering
\subfigure[]{
        \includegraphics[height=6cm]
                {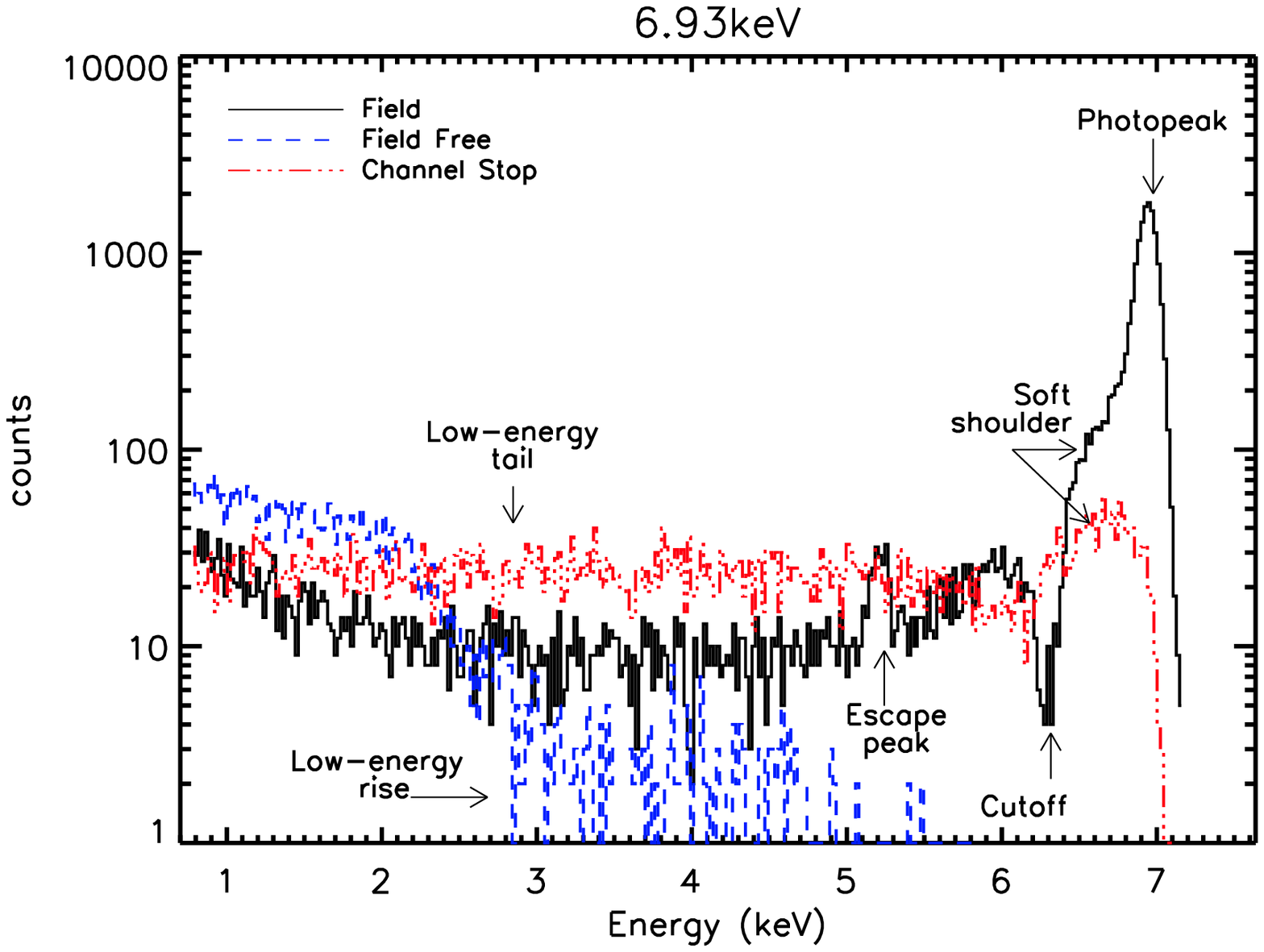}
\label{co}%
}%
\qquad
\subfigure[]{
        \includegraphics[height=6cm]
                {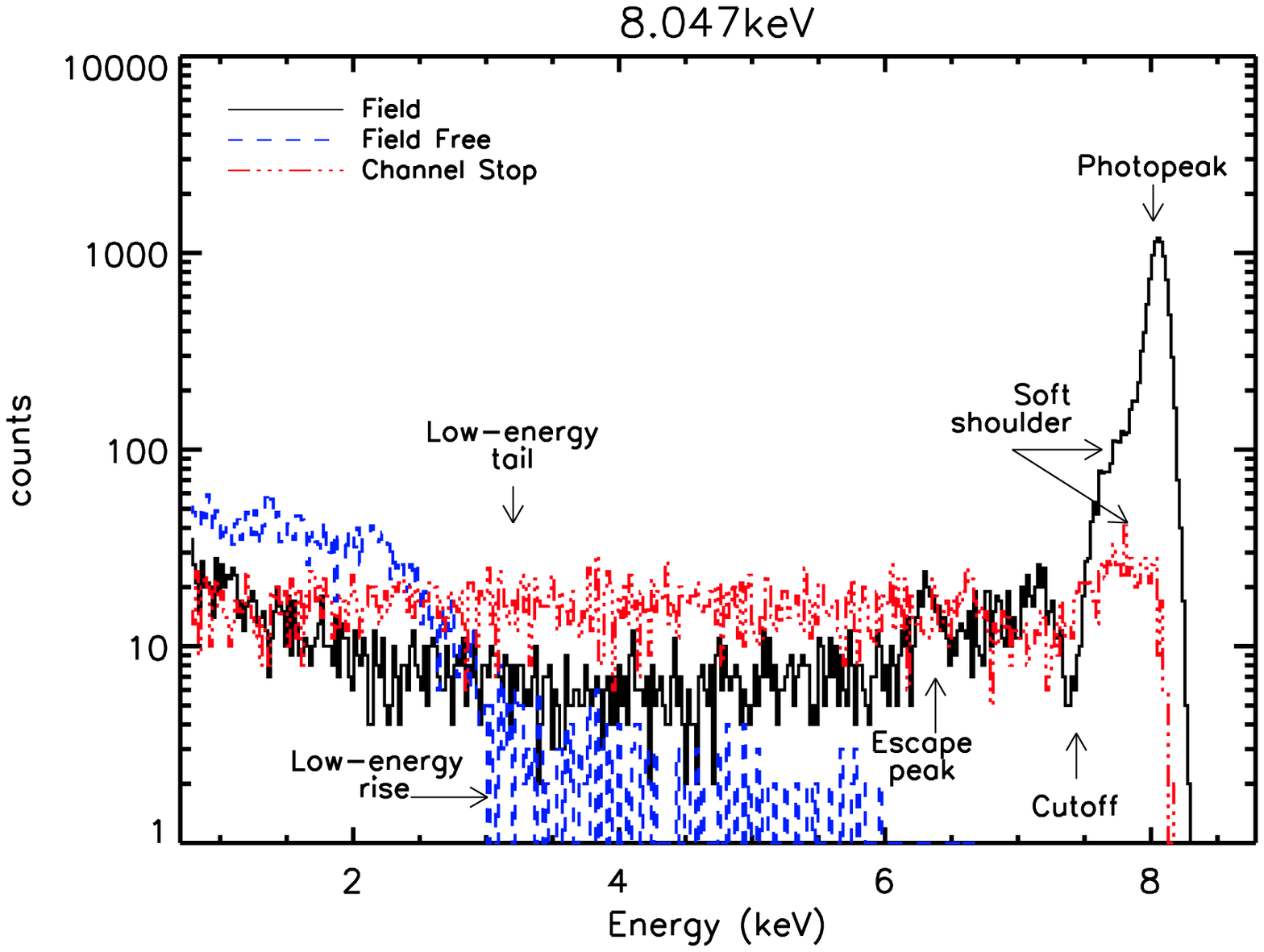}
\label{cu}%
}%
\caption[SRF of SCD with components - simulation results (6.93 keV, 8.047 keV)]{Simulation results showing different components of SRF for
different mono-energetic photons arising from interactions at different
layers of SCD: (a) 6.93 keV (b) 8.047 keV.}
\end{figure}

  Channel stop interactions contribute partly to the soft
  shoulder component (near the photopeak) and also to the low-energy tail in the
  SRF as shown in Figs. \ref{ti}, \ref{cr}, \ref{co}, and \ref{cu}.

  \subsection{Comparison of SRF simulation Vs SRF data}
We demonstrate the performance of the model by comparing the simulated SRF
with the SRF derived from C1XS ground calibration. It is clear that all the observed features arise from
interactions at different layers of the detector, coupled with the threshold
values and event selection logic. An overplot of simulation and calibration data that are normalized to total
events for different energies is shown in Figs. \ref{tiop}, \ref{crop}, \ref{coop} and \ref{cuop}.

\subsection{Statistical tests}
To establish the robustness of the algorithm, the following standard
statistical tests are performed on the data sets: SRF derived from simulation and
calibration at different energies. A simple Student`s T-test is performed to
verify the null hypothesis that the simulation and calibration data exhibit same
means which are drawn from a population with same variance.  The computed T-statistic
values and significance are given in Table \ref{stests}. The very low T-statistic values
with high significance clearly show that the SRF of model and calibration have
same mean values.

A chi-square (${\chi}^2$) test is often used to determine whether two binned data sets
belong to the same parent distribution. It computes the differences between
the binned data sets. The chi-square statistic computed for our data sets, the
number of degrees of freedom (${\nu}$), and the corresponding chi-square probability
are given in Table \ref{stests}. A low ${\chi}^2$ value with high degrees of freedom (${\nu}$) yields a high probability that clearly indicates that the SRF derived from model and calibration data agree very closely.

\begin{figure}%
\centering
\subfigure[]{
        \includegraphics[height=6cm]
                {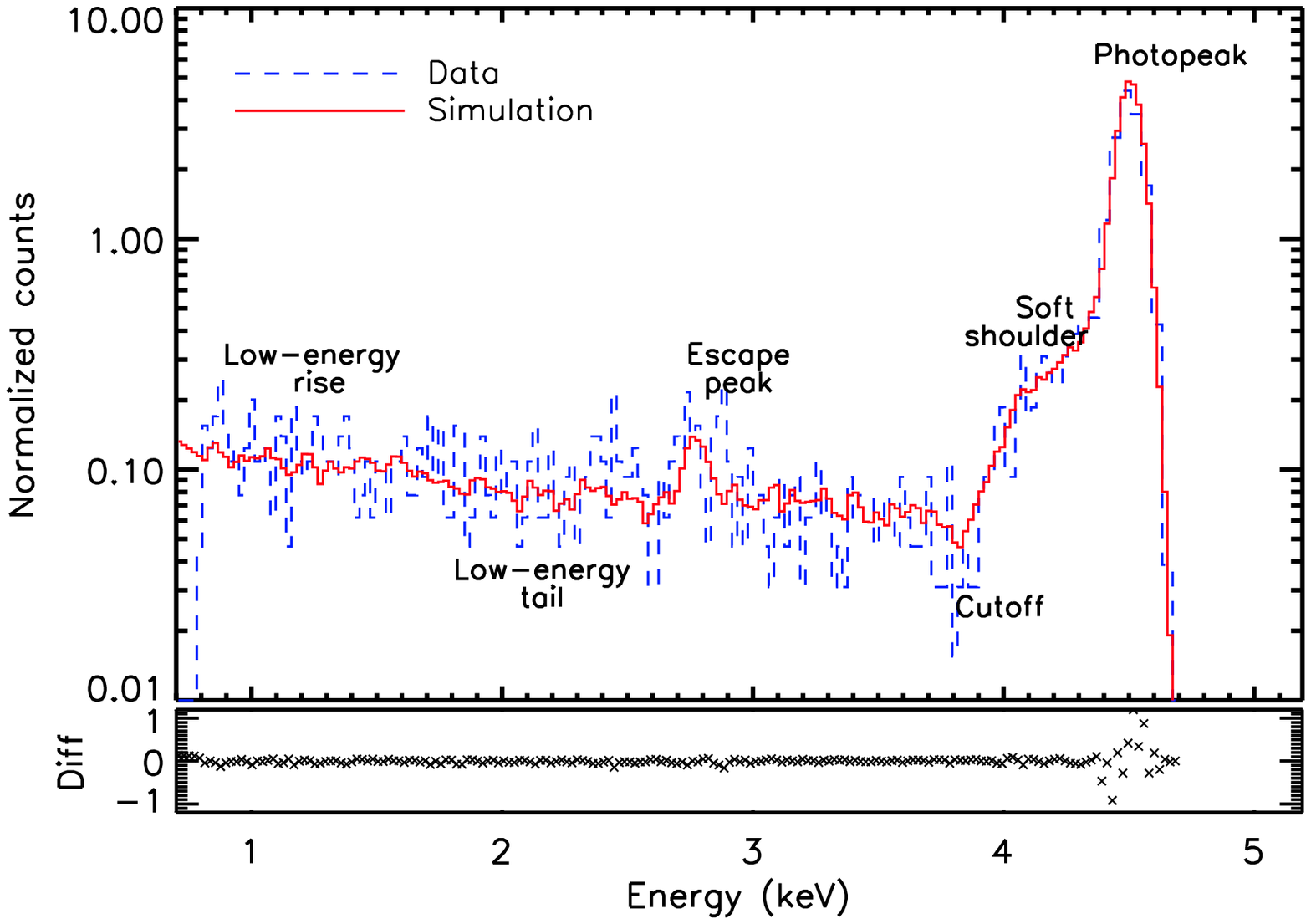}
\label{tiop}%
}%
\qquad
\subfigure[]{
        \includegraphics[height=6cm]
                {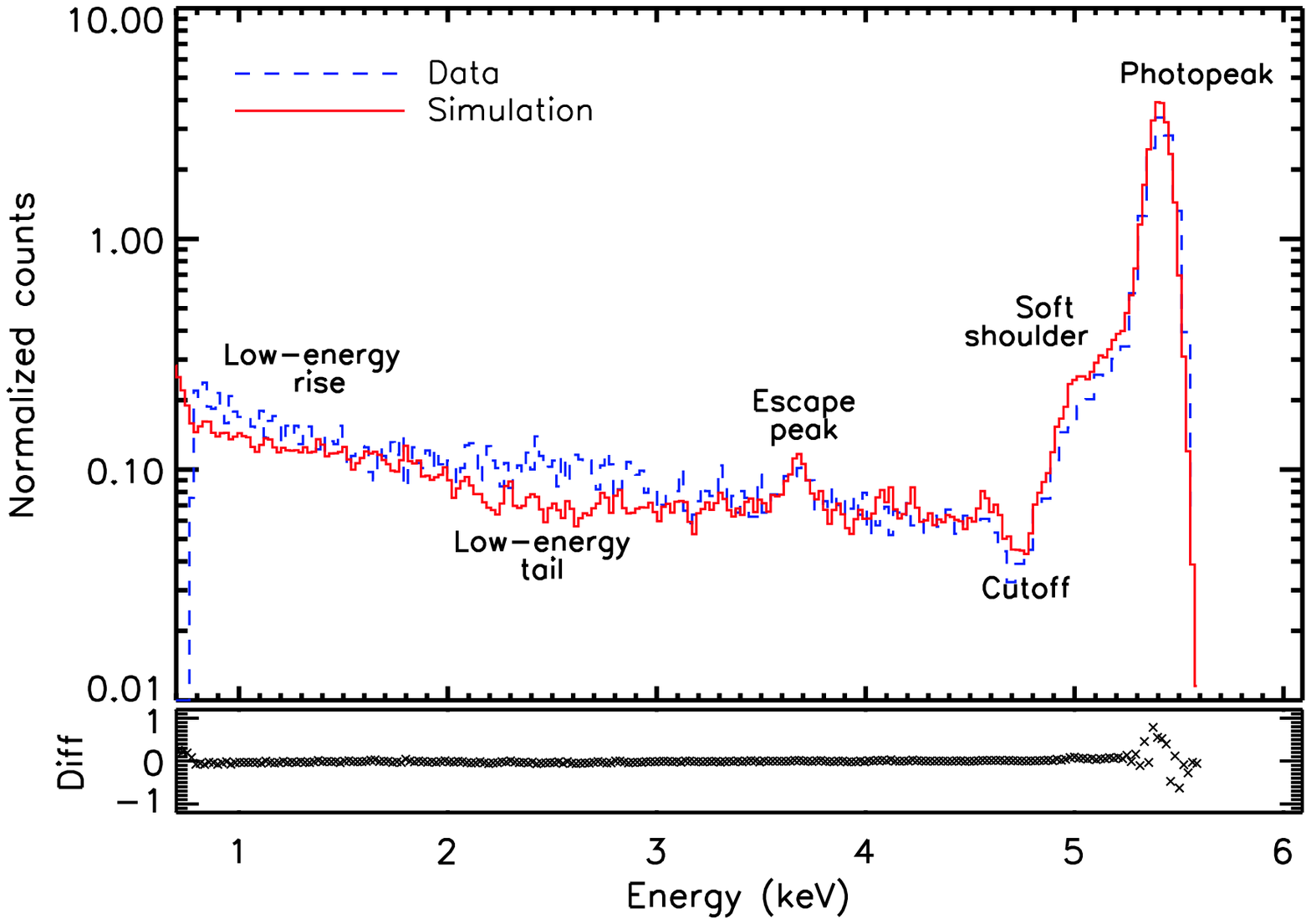}
\label{crop}%
}%
\caption[Overplot of SRF obtained from simulation and C1XS ground calibration
(4.510 keV, 5.414 keV)]{Comparison of simulated SRF and the observed SRF from C1XS ground
calibration: (a) 4.510 keV (b) 5.414 keV. The bottom panel shows the difference between two SRFs. }
\end{figure}

\begin{figure}%
\centering
\subfigure[]{
        \includegraphics[height=6cm]
                {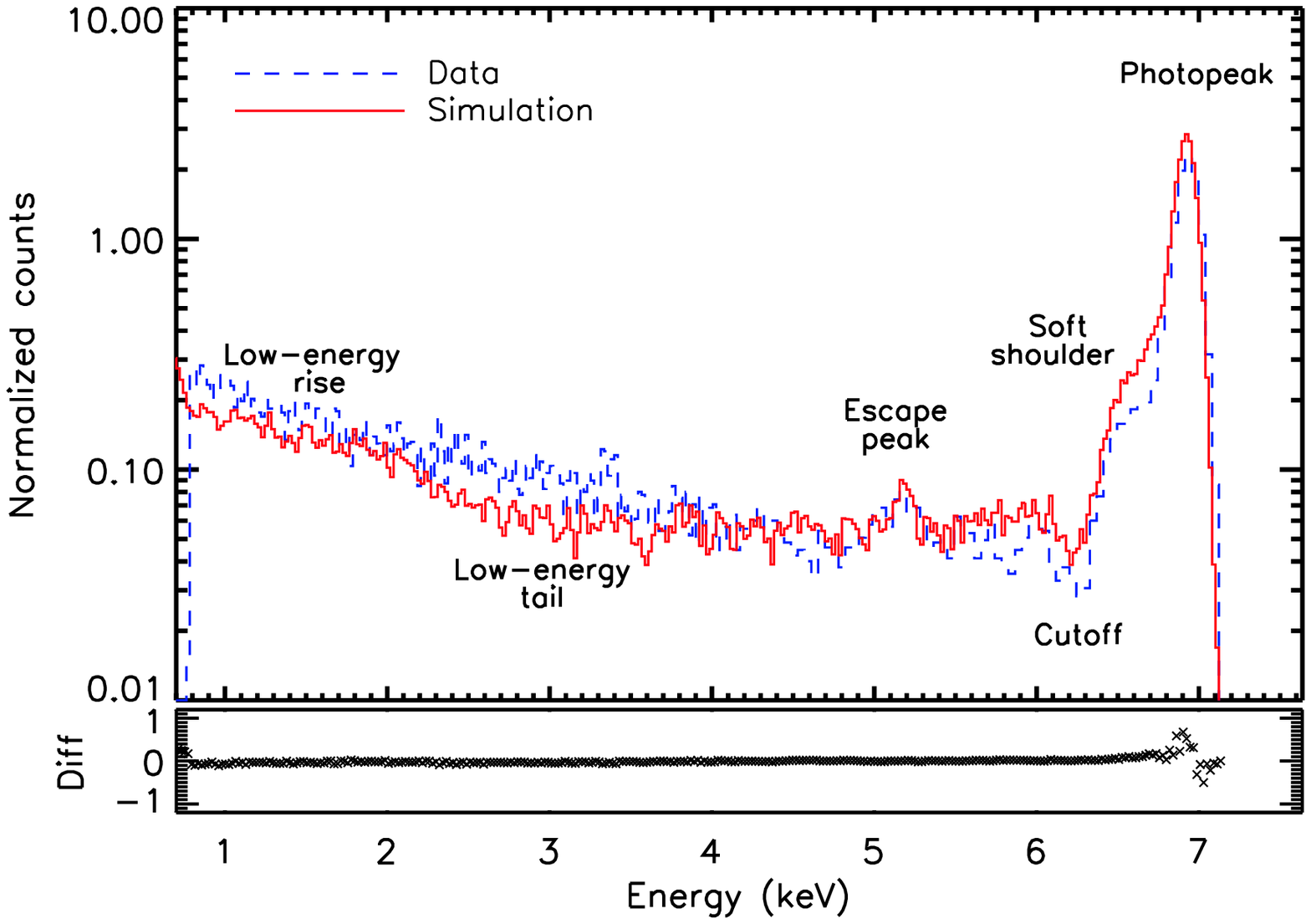}
\label{coop}%
}%
\qquad
\subfigure[]{
        \includegraphics[height=6cm]
                {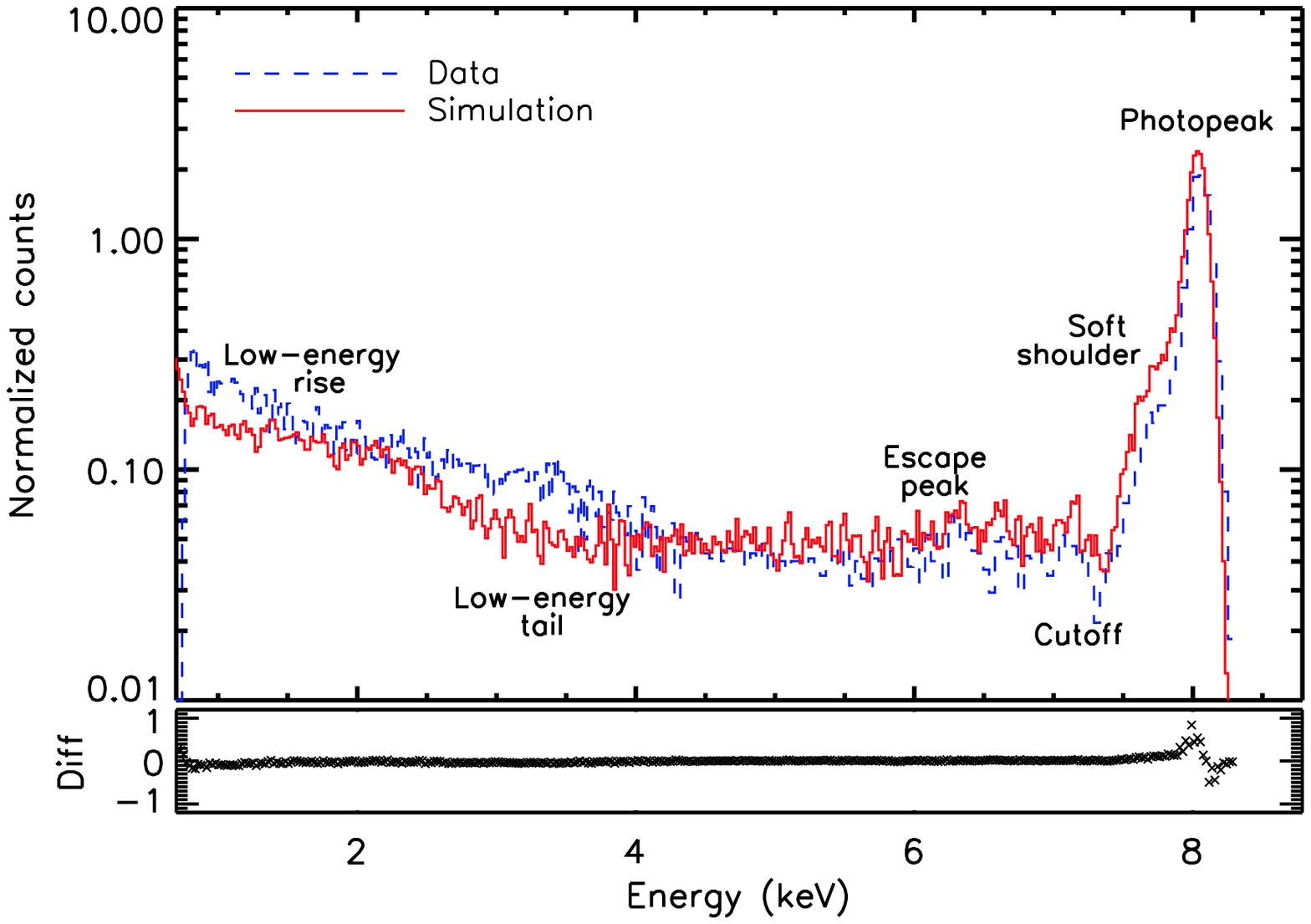}
\label{cuop}%
}%
\caption[Overplot of SRF obtained from simulation and C1XS ground calibration
(6.93 keV, 8.047 keV)]{Comparison of simulated SRF and the observed SRF from
C1XS ground calibration: (a) 6.93 keV (b) 8.047 keV.  The bottom panel shows the difference between two SRFs.}
\end{figure}

\begin{figure}
  \centering
  \includegraphics[width=10cm]{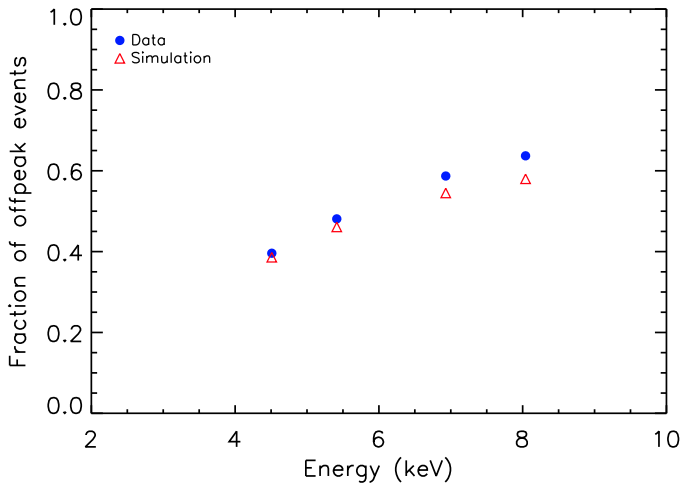}
  \caption{Comparison of fraction of off-peak events as a function of photon
  energy between simulation (triangles) and calibration data (filled circles).}
\label{offpeak}%
  \end{figure}

Our simulation results predict all the major features seen in the experimentally
derived SRF. The difference between data and model are represented as residuals
in the bottom panel of each figure (Figs. \ref{tiop} - \ref{cuop}). High energy
X-ray photons penetrate deep inside the detector before interaction, hence
additional interactions in the field-free zone causing more split events. The
expected trend toward decreases in the photopeak and increases in off-peak events with
increased X-ray energy is clearly seen.  The variation in the fraction of
off-peak events at different energies obtained from calibration data and
simulation are plotted in Fig. \ref{offpeak}, and a close match between the two
clearly shows that the model represents the calibration data fairly well. It is
noted that deviations are observed in predicting soft shoulder and low-energy
tail components at higher X-ray energies. We attribute this discrepancy to
incomplete modeling of the energy dependence in the channel stop interactions.

  \section{Conclusion}
  \label{conc}
  To conclude, we have successfully demonstrated the capability of our charge
  transport model in simulating the SRF of epitaxial SCDs used in C1XS.  Comparing the simulation results to the available ground calibration data clearly demonstrates that the SRF can be determined at different energies. The validation of simulation results with C1XS ground calibration data have
  provided physical reasoning for all the observed spectral features and also its energy
  dependence. Simulation results show a systematic underestimation of the fraction of
  offpeak events with an increase in photon energy. This discrepancy can be
  mitigated by including the  energy dependence of channel-stop tuning
  parameters, and this would require further detailed experiments. Also the interaction within the
  dead layer and substrate are considered to be lost, which would need to be
  included for very low energy and high energy X-ray photons.

\begin{table}
\caption{Simulated SRF Vs Observed SRF - Statistical tests results }             
\label{stests}      
\centering                          
\begin{tabular}{|c|c|c|c|c|}        
\hline                 
 Photon  & \multicolumn{2}{c|}{Student's T-test}  & \multicolumn{2}{c|} {Chi-square test}\\
\cline{2-5}    
 energy (keV) &T-statistic      & Significance & ${\chi}$$^{2}$ statistic & Significance\\
\hline                        
4.510 & -0.021&0.983&2.67&1.0\\
5.414 & -0.075&0.941&2.48&1.0\\
6.93  & -0.109&0.913&3.49&1.0\\
8.047 & -0.123&0.902&4.24&1.0\\
\hline                                   
\end{tabular}
\end{table}

\bibliographystyle{aa}
\bibliography{sample}  
\end{document}